\documentclass[pre,superscriptaddress,showpacs,eqsecnum,twocolumn]{revtex4}
\usepackage{newlfont}
\usepackage{amssymb}
\usepackage{amsfonts}
\usepackage{amsmath}
\usepackage{graphicx}
\usepackage{rotating}
\usepackage{bm}
\usepackage[scanall]{psfrag}
\begin{document}
\title{Superfluidity of the BEC at finite temperature.}
\author{{\L}ukasz Zawitkowski}
\affiliation{Center for Theoretical Physics, Polish Academy of Sciences, Aleja Lotnik{\'o}w 32/46, 02-668 Warsaw, Poland}
\author{Mariusz Gajda}
\affiliation{Institute of Physics, Polish Academy of Sciences, Aleja Lotnik{\'o}w 32/46, 02-668 Warsaw, Poland}
\affiliation{Faculty of Mathematics and Sciences, Cardinal Stefan Wyszy\'nski University, Warsaw, Poland}
\author{Kazimierz Rz{\c a}{\.z}ewski}
\affiliation{Center for Theoretical Physics, Polish Academy of Sciences, Aleja Lotnik{\'o}w 32/46, 02-668 Warsaw, Poland}
\affiliation{Faculty of Mathematics and Sciences, Cardinal Stefan Wyszy\'nski University, Warsaw, Poland}

\date{\today}

\begin{abstract}
We use the classical fields approximation to study a translational flow of the condensate with respect to the thermal cloud
in a weakly interacting Bose gas.
We study both, subcritical and supercritical relative velocity cases and analyze in detail a state of stationary flow which is reached in the dynamics. 
This state corresponds to the thermal equilibrium, which is characterized by the relative velocity of the condensate and the thermal cloud.
The superfluidity manifests itself in the existence of many thermal equilibria varying in 
(the value of this velocity) the relative velocity between the condensate and the thermal cloud.
We pay a particular attention to excitation spectra in a phonon as well as in a particle regime. 
Finally, we introduce a measure of the amount of the superfluid fraction in a weakly interacting Bose gas,
allowing for the precise distinction between the superfluid and the condensed fractions in a single and consistent framework.

\end{abstract}

\maketitle

\section{Introduction}
The tendency of liquid helium-4 cooled below 2.19 K to flow without any apparent friction is known for decades 
\cite{kePRAA_HistorySuperfluidity,kNATURE_HistorySuperfluidity,amNATURE_HistorySuperfluidity}. 
This superfluid behavior is commonly associated with the phenomenon of a Bose-Einstein condensation \cite{dgpsRMP1999}.  
The connection has been mostly based on the similarities in the nature of both phenomena,
as both are a large-scale manifestations of the quantum nature.
However, the detailed theoretical investigation of superfluidity and condensation in liquid helium \cite{cRMP1995} is a great challenge,
as it is a strongly interacting quantum system and cannot be effectively described with a mean-field or perturbative approaches.
The present-date understanding of superfluidity comes from the paper of L. Landau \cite{lJPMoscow1941_superfluidity}.

Experimental achievement of Bose-Einstein condensates in dilute trapped atomic gases \cite{BEC1,BEC2,BEC3} 
gives a unique possibility to study their superfluid properties under condition of very weak interactions. 
It has opened an opportunity for revisiting the concept of superfluidity.
In this context the issue of the atomic Bose-Einstein condensate under rotation has attracted great interest \cite{fsJPC2001}.
It has been demonstrated experimentally that Bose-Einstein condensate circulates around quantized vortex lines. 
In experiments with gasous BEC the existence of these vortices and the properties of scissor modes 
\cite{mcwdPRL2000_vorticesBEC,mahhwcPRL1999_vorticesBEC,hhhmfPRL2002_vorticesBEC,rkodkhkPRL1999_experimentalBEC_criticalvelocity,mhahhfPRL2000_scissors} 
are considered the main manifestations of superfluidity.
On the other hand, a little is known about superfluidity in a translational motion of the atomic condensate 
\cite{apPRA2004,pPRA2002_superfluidityinatomlaser1d,lclcPRL2000_superfluidityabovecriticalvelocity}.
There are still no experiments that would correspond to the demonstration of frictionless non-rotary flow,
which is the most intuitive manifestation of this phenomenon.

Dilute atomic Bose gases are simpler to describe than liquid helium, with many succesful implementations of mean-field theories.
It is a purpose of this paper to present a simple model of superfluidity in a weakly interacting Bose gas using the classical fields approximation.
The classical fields approximation is the non-perturbative mean-field model of a Bose-Einstein condensate at finite temperatures.
It turned out to be very successful in describing dynamical and 
thermodynamic properties of atomic condensates and their excitation spectra \cite{bpgrJPB2004,lscPRL2004,slcPRL2001,ggrPRA2002,dmbPRL2001,dmbPRA2002}. 
It allows a description of the condensate and a thermal cloud treating both components on equal footing without any arbitrary splitting of the system 
into two parts. 

In Section \ref{The Method} we briefly describe the classical fields approximation.
For ease of calculations we restrict ourselves to a simplified geometry of 3D box with periodic boundary conditions. 
Then we prepare initial states which will allow us to study the relative motion of the condensate and the thermal cloud.

In Section \ref{Numerical Results} we present numerical results.
We find that the flow is indeed frictionless below the Landau critical velocity $c$ 
and that the stationary state depends on yet another control parameter - namely the relative velocity between the condensate and the thermal cloud.
We believe that in this paper we are extracting for the first time the basic properties of the superfluid flow
in direct numerical simulation of the dynamics.
We obtain the excitation spectra of such a superfluid system and observe the equipartition of energies in quasiparticle modes,
which has been previously found in case of a stationary BEC \cite{dmbPRA2002,bpgrJPB2004}.
We also notice pulling of Bogoliubov quasiparticles by the condensate during the thermalization period.

In Section \ref{Analitical Results} we analyze the properties of Bogoliubov quasiparticles.
We derive analitically excitation spectra of the system, obtaining results that agree with our numerical simulations. 
We analyze the transformation of quasiparticle momentum under Galilean transformations and find that it behaves as a particle of mass $m$, 
similarly to atoms in the system and unlike acoustic waves.
This allows us to directly relate superfluid and condensed fractions,
what produces surprisingly simple formula in the case of 3D box.
We base our reasoning on the observation that Bogoliubov quasiparticles are the normal modes of the system \cite{bpgrJPB2004}
and we associate with them the normal (non-superfluid) fraction, similarly to Landau.
In Section \ref{Conclusions} we conclude.

\section{The Method}\label{The Method}

We consider a weakly-interacting Bose gas consisting of N atoms confined in a 3D box potential with periodic boundary conditions.
The atoms interact via a contact potential $V(r-r^\prime)=\frac{4\pi\hbar^2a}{m}\delta^{(3)}(\textbf{r}-\textbf{r}^\prime)$, where $a$ is the s-wave scattering length.
They can be described by the second-quantized Hamiltonian:
\begin{equation}\label{Hamiltonian}
  H=\int_{L^3} d^3x \ (\hat{\Psi}^\dagger \frac{\hat{p}^2}{2m} \hat{\Psi})+\frac{2\pi^2\hbar^2a}{m} 
\int_{L^3} d^3x \ (\hat{\Psi}^\dagger \hat{\Psi}^\dagger \hat{\Psi}\hat{\Psi}),
\end{equation}
where $L$ is the size of the box, $m$ is the mass of an atom and
$\hat{\Psi}$ is a bosonic field operator satisfying equal time bosonic commutation relation 
$[\hat{\Psi}(r,t), \hat{\Psi}^{\dagger}(r^\prime,t)]=\delta^{(3)}(\textbf{r}-\textbf{r}^\prime)$.
The corresponding Heisenberg equation for $\hat{\Psi}$ is of the form:
\begin{equation}\label{Heisenberg}
  i\hbar\partial_t \hat{\Psi}=-\frac{\hbar^2\Delta}{2m}\ \hat{\Psi}+\frac{4\pi\hbar^2a}{m} \hat{\Psi}^\dagger \hat{\Psi}\ \hat{\Psi}.\\
\end{equation}
We expand the field operator in the basis of plane waves: 
$\hat{\Psi}=\frac{1}{\sqrt{L^3}} \sum_{\textbf{k}} e^{2\pi i \textbf{kr}/L} \hat{a}_{\textbf{k}}(t)$,
where annihilation operators $\hat{a}_{\textbf{k}}$ destroy particle in mode $\textbf{k}$
and satisfy a commutation relation $[\hat{a}_{\textbf{k}},\ \hat{a}^{\dagger}_{\textbf{k}^\prime}]=\delta_{\textbf{k},\textbf{k}^\prime}$.

Now we apply the classical fields approximation \cite{zbgrPRA2004, ggrPRA2002,ggrOEx2001,dmbPRL2001,slcPRL2001}.
For all modes that are occupied by sufficiently large number of atoms we subtitute annihilation operators with c-numbers: 
$\hat{a}_{\textbf{k}} \rightarrow \sqrt{N} \alpha_{\textbf{k}}$.
After neglecting all remaining operator terms we obtain a set of equations:
\begin{equation}\label{PlaneGPE}
  \partial_t \alpha_{\textbf{k}}(t)=-i\frac{2\pi^2\hbar}{mL^2}\ \textbf{k}^
2\ \alpha_{\textbf{k}}(t)-i\frac{4\pi\hbar aN}{mL^3}
 \sum_{\textbf{k}_1,\ \textbf{k}_2} 
  \alpha^*_{\textbf{k}_1}\alpha_{\textbf{k}_2}\alpha_{\textbf{k}+\textbf{k}_1-\textbf{k}_2}
\end{equation}

We choose wavectors of "classical" modes $\alpha_{\textbf{k}}$ to span a grid of size 32x32x32.
They compose the normalized mean-field wave function 
$\psi=\frac{1}{\sqrt{L^3}} \sum_{\textbf{k}} e^{2\pi i \textbf{kr}/L} \alpha_{\textbf{k}}(t)$.
This wavefunction obeys the Gross-Pitaevskii equation on a spatial grid:
\begin{equation}\label{GPE}
  i\hbar \partial_t \psi= -\frac{\hbar^2\Delta}{2m} \psi+ L^3 \epsilon g |\psi|^2 \psi,
\end{equation}
where $g=a N/\pi L$ is the interaction strength and  
 $\epsilon=4\pi^2\hbar^2/mL^2$ is our unit of energy 
(the corresponding unit of time is $\hbar/\epsilon$).

The wavefunction $\psi$ represents here both the condensate and the thermal cloud,
contrary to the typical interpretation of the GP equation
where $\psi$ is just the condensate wavefunction.
A measurement, which introduces a coarse-graining is essential for distinguishing between these two fractions. 
Populations of various modes can be extracted from diagonalization of the time-averaged density matrix \cite{Penrose, ggrPRA2002}
and the dominant eigenvalue represents the condensed fraction. 
The time-averaging destroys the coherence between the modes and leads to a mixed state out of a pure state $|\psi><\psi|$. 
The wave-function $\psi(t)$ can be interpreted as a single experimental realization of evolving system.

To obtain a stationary solution corresponding to non-zero temperature we start with the ground state wave-function and randomize its phase.
The system evolves towards a thermal equilibrium, which depends only on the total energy per particle $E$,
the interaction strength $g$ and the length of the box $L$ \cite{ggrPRA2002}.
The condensate appears in the zero momentum mode and atoms in the remaining modes compose a thermal cloud.
The dependence of the resulting state on the interaction strength $g=aN/\pi L$ instead of $a$ and $N$ separately results in an ambiguity.
In order to assign the temperature $T$ to our system in a way independent of the choice of the numerical grid 
we use a scheme developed in \cite{zbgrPRA2004};
we assign temperature to temperature of the ideal gas that has the same condensate fraction.

In the considered case of the box potential eigenmodes of the stationary single-particle density matrix 
can be only composed of modes with the same $|\textbf{k}|$, due to symmetry constrains.
Thus the plane waves set a natural basis. 
As it turns out, modes $\textbf{k}$ combine with modes $\textbf{-k}$ to create Bogoliubov quasiparticles of amplitude $\delta_{\textbf{k}}$,
which evolve with a single dominant frequency
(the ralation between $\delta_{\textbf{k}}$ and $\alpha_{\textbf{k}}$ will be derived in Section \ref{Analitical Results}).
Bogoliubov quasiparticles are the normal modes of the system \cite{bpgrJPB2004};
for high momenta, when the kinetic energy is much larger than the interaction energy,
they become indistinguishable from atomic modes $\alpha_{\textbf{k}}$. 

In a stationary case the central energies of quasiparticles are given by \cite{bpgrJPB2004}:
\begin{equation}\label{excitations_stationary}
  \epsilon_{\textbf{k}}=\mu+\epsilon \sqrt{\omega(\textbf{k})^2-(gn_c)^2},
\end{equation}
where $\omega(\textbf{k})=\frac{\textbf{k}^2}{2}+gn_c$ is introduced for simplicity and
$n_c$ is the condensate fraction. 
$\mu$ is the frequency of the condensate mode and has an interpretation of the chemical potential.
Similarly to the case of He-II they have a phonon-like spectrum at low momenta (see Fig. 5 in \cite{bpgrJPB2004}).
In the classical fields approximation quasiparticles obey the equipartition of energies \cite{bpgrJPB2004},
suggesting that the system is in a thermal equilibrium and not just in a steady state.
Note, however, that the condensate itself is not populated according to the equipartition.
For parameters considered in our paper the Bose statistics leads to the equipartition for the considered Bogoliubov quasiparticles,
suggesting that thermodynamics should be studied in quasiparticle and not atomic modes.

The solutions discussed here represent the situation in which the condensate does not move with respect to the thermal cloud.
In order to study the possible superfluid flow we need to put the condensate in motion.
In our numerical simulations we take a stationary state described above
and apply a momentum kick to the condensate.
The remaining atoms are pushed in the opposite direction, so that the total momentum of the system is close to zero
\footnote{
This is neccessary due to a numerical constraint imposed by the momentum cut-off, 
which introduces artificial umklapp processes near the border of the numerical grid.
Only when the total momentum of the system is zero these unphysical processes do not alter the dynamics.}.
We can also apply the momentum kick to any arbitrary number of modes, 
in particular to the condesnate along with surrounding modes which merge into quasiparticles (phonon modes).

In our simulations we take stationary solutions for ca. $60000$ atoms and the scattering length of $18\ nm$, confined in a box of lenght $5\ \mu m$, 
with the condensate populations of $65\%$ and $8\%$. The temperatures are $580\ nK$ and $1\ \mu K$, respectively.
We apply the procedure described above, controlling the initial relative velocity $v$ of the two fractions, by placing the condensate in a chosen mode $\textbf{k}_c$.
The resulting states are composed of the condensate (possibly with some surrounding atomic modes) and the remaining thermal cloud
flowing in opposite directions.
These states are arbitrarily prepared and no longer represent an equilibrium of the system - they should experience some thermalization.

Can we observe any frictionless (superfluid) flow between the two fractions? 

\section{Numerical Results}\label{Numerical Results}

\begin{figure}[t]
\includegraphics[width=0.5\textwidth]{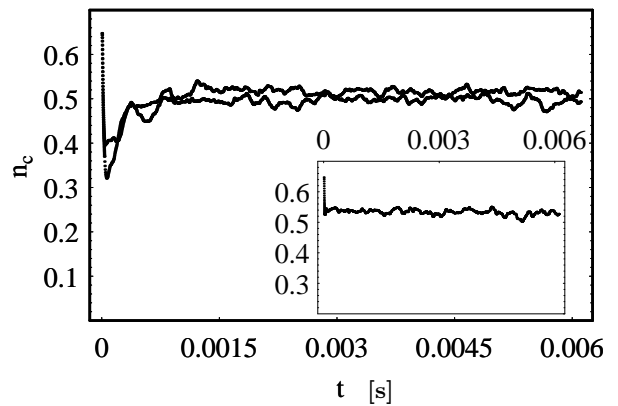}
\put(-103,7){\mbox{$\textbf{[s]}$}}
\caption{
Fraction of the condensate versus time for the case of superfluid flow with initial velocities between the thermal cloud and the condensate
equal $v=0.48c$ and $v=0.82c$. The equilibrium condensate fraction is $51.6\%$ and $49.6\%$, respectively.
The initial states were prepared by pushing the condensate without phonon modes. 
In this case dragging of phonons has occured.
The inset shows a situation where phonon modes have been initially pushed with the condensate and $v=0.79c$.
The resulting condensate fraction is $53.4\%$ in this case.
} \label{superflow}
\end{figure}

The steady state depends on the initial relative velocity $v$ between the condensate and the thermal cloud.
There exists a critical velocity $c=\frac{2 \pi \hbar}{mL} \sqrt{gn_c}$ which separates two regimes;
$n_c$ is the new equilibrium condensate fraction.
For a system with $65\%$ ($8\%$) of the condensate initially
the resulting $n_c$ is approximatelly $50\%$ ($1.5\%$) for velocities lower than $c$.
For relative velocities v below the critical velocity c the condensate remains flowing with respect to the thermal cloud.
No drag due to thermal atoms is experienced and the flow is superfluid (Fig. \ref{superflow}).
Note the initial thermalization.
We have observed that during this transient period the phonon modes are being dragged along with the condensate. 
Initially pushing low-excited modes along with the condensate reduces this time and increases the equilibrium condensate fraction
(see inset in Fig. \ref{superflow}).

\begin{figure}[t]
\includegraphics[width=0.5\textwidth]{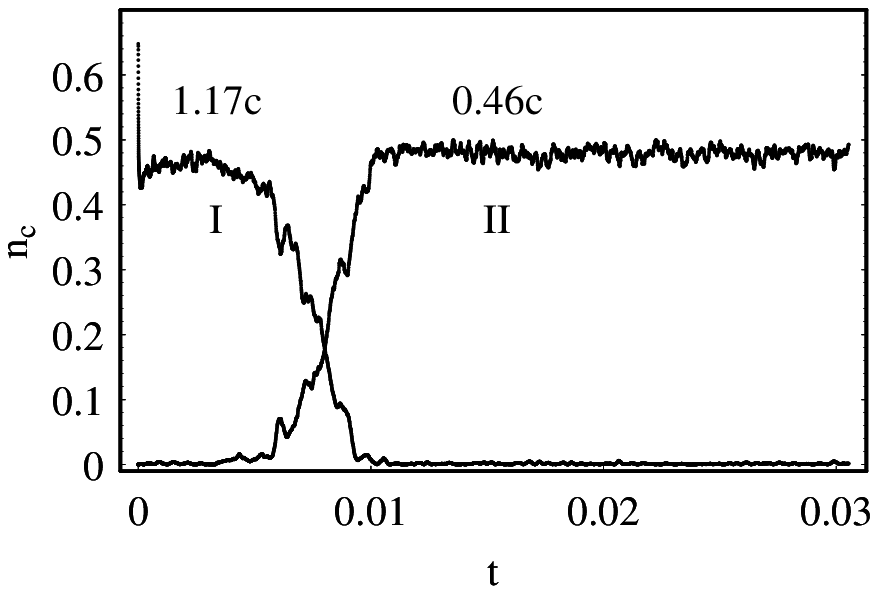}
\put(-105,8){\mbox{$\textbf{[s]}$}}
\put(-50,58){\mbox{\large{(a)}}}

\includegraphics[width=0.5\textwidth]{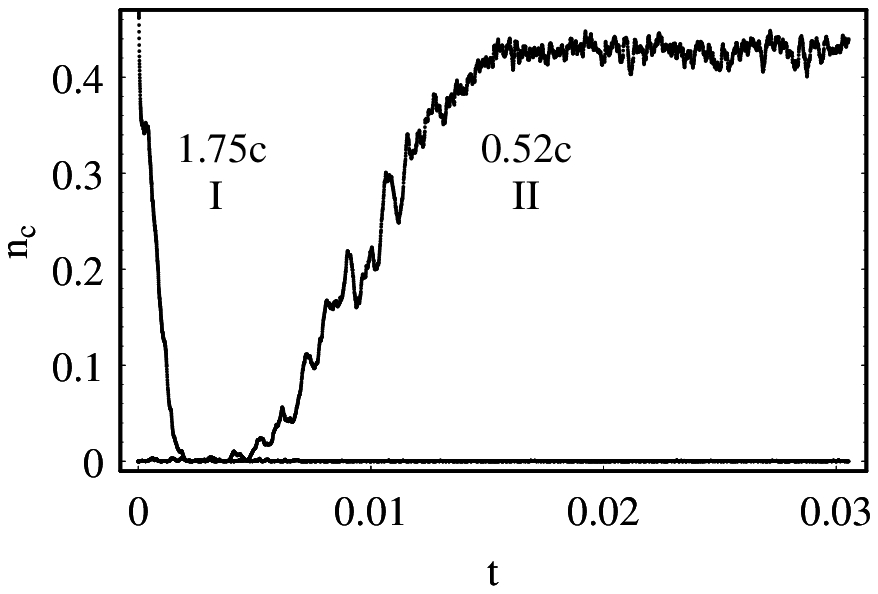}
\put(-105,8){\mbox{$\textbf{[s]}$}}
\put(-50,58){\mbox{\large{(b)}}}
\caption{
Population of the condensate versus time for a supercritical flow. 
The starting relative velocity of the condensate and the thermal cloud is $v=1.17c$ (a) and $v=1.75c$ (b). 
The mode initially occupied by the condensate becomes depleted due to friction with the thermal cloud (I) and 
the condensate reappears in a slower moving mode (II).
Note, that in (a) the condensate is directly transferred to the neighbouring momentum state, 
where it flows with a stationary relative velocity $v=0.46c$ with respect to the thermal cloud.
On the contrary in (b), where the initial velocity is very high, the flow is much more complicated.
The condensate population smears into multiple modes to finally agregate in a mode 
whose relative velocity equals $v=0.52c$.
In the transient time many succeeding slips may occur, as the condensate slows down gradually.
As a result of these slips the stationary relative velocity of the condensate and the thermal cloud is subcritical. 
The equilibrium condensate population is $47.9\%$ in (a) and $42.9\%$ in (b).
} \label{normalflow}
\end{figure}

\begin{figure}[t]
\includegraphics[width=0.5\textwidth]{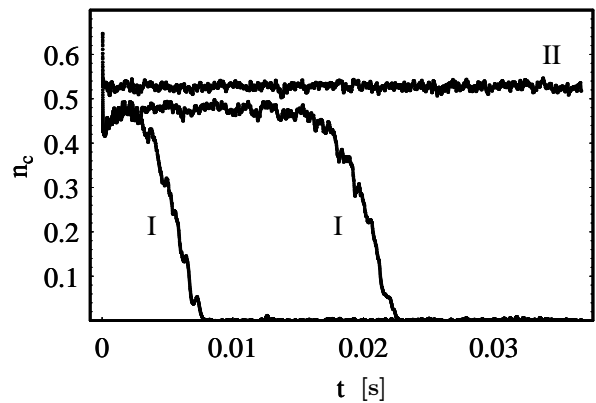}
\put(-103,8){\mbox{$\textbf{[s]}$}}
\caption{
Fraction of the condensate versus time for a flow with the critical velocity.
(I) The condensate has been pushed without phonon modes and the initial relative velocity between the condensate and the thermal cloud is $v=c$.
The two plots represent different realizations of the same setup. 
Slips occur at different time and show that the system is unstable in the critical region.
(II) Phonon modes have been initially pushed with the condensate and stabilize the flow, which is now superfluid. 
The relative velocity between the condensate and the thermal cloud is now $v=0.98\ c$ due to a slightly greater condensate population.
} \label{criticalflow}
\end{figure}

Above the critical velocity the condensate experiences a drag from thermal atoms and slows down.
The condensate population slips to a slower moving mode (Fig. \ref{normalflow}).
Close to the critical velocity the system is unstable.
Slips will occur at different times, 
depending on the particular realization of the initial conditions (Fig. \ref{criticalflow}).
Bundling phonon modes with the condensate results in stabilizing the near-critical flow.
This is due to an increase in the resulting equilibrium condensate fraction which shifts 
the critical velocity slightly upwards.

\begin{figure}[t]
\includegraphics[width=0.5\textwidth]{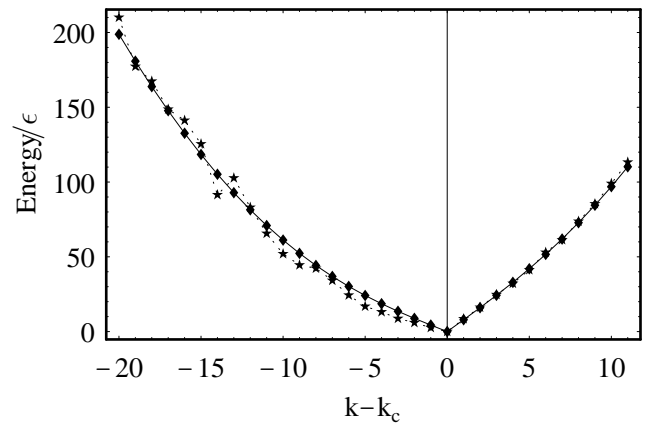}
\put(-245 ,108){\begin{sideways}\mbox{\large{\textbf{$/\epsilon$}}}\end{sideways}}
\caption{The excitation spectra of quasiparticles in a direction along the momentum kick. 
Numerical results (stars) were obtained for the case of the initial condensate fraction $n_c=0.65$ pushed together with the phonon modes with
the initial velocity relative to the thermal cloud equal $v=0.82 c$.
Dots depict a fit with Formula \ref{excitations_eqaution}.
The equilibrium condensate fraction is $n_c=0.534$. 
Spectra taken in directions different than presented experience smaller tilting in agreement with Formula \ref{excitations_eqaution}.
} \label{excitation_spectra}
\end{figure}

\begin{figure}[b]
\includegraphics[width=0.5\textwidth]{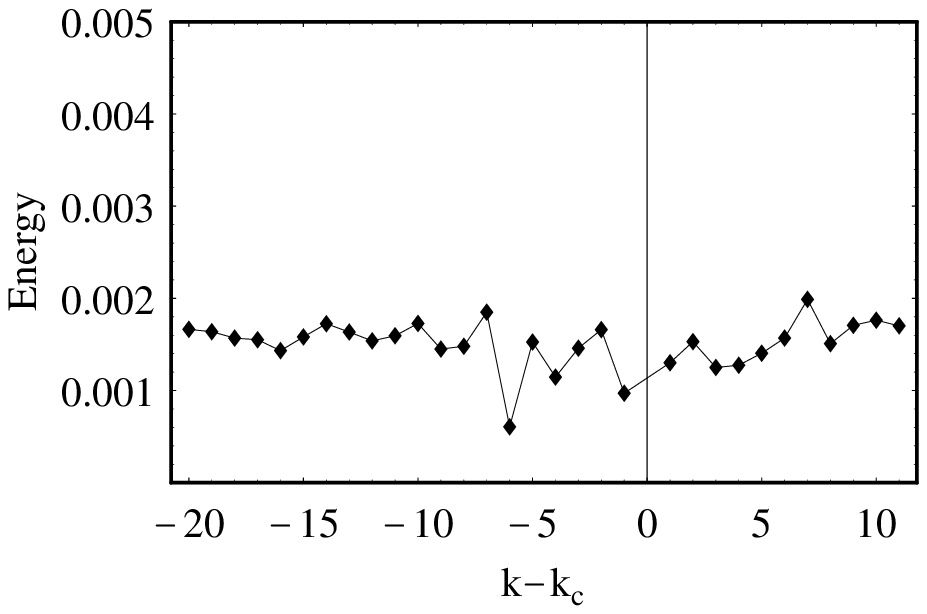}
\put(-243 ,105){\begin{sideways}\mbox{\large{\textbf{$/\epsilon$}}}\end{sideways}}
\put(-180 ,125){\mbox{\large{(a)}}}

\includegraphics[width=0.5\textwidth]{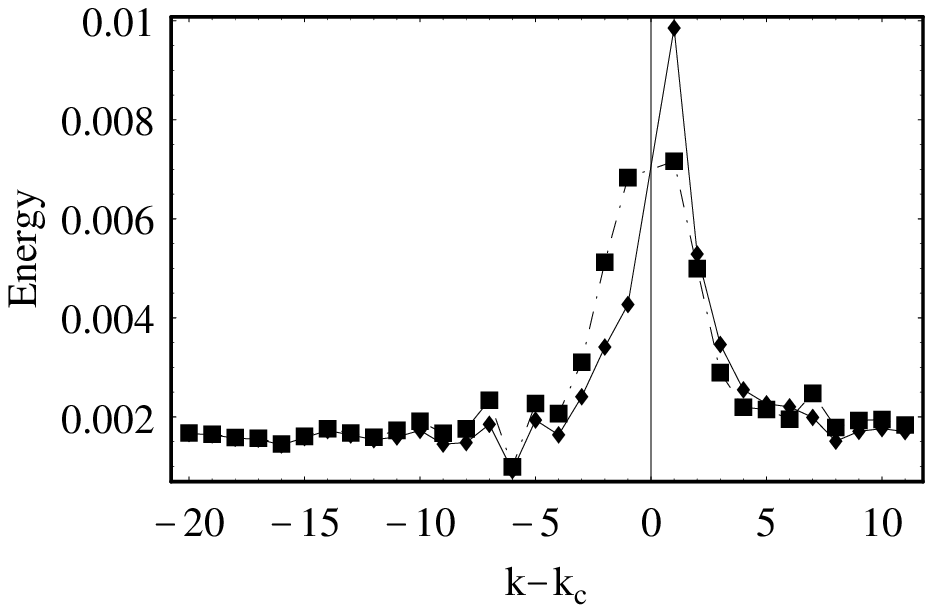}
\put(-243 ,105){\begin{sideways}\mbox{\large{\textbf{$/\epsilon$}}}\end{sideways}}
\put(-180 ,125){\mbox{\large{(b)}}}
\caption{Energies localized in quasiparticle modes (a) and atomic modes (b) versus their momenta for parameters as in Fig. \ref{excitation_spectra}.
Squares in (b) are analitical predictions for energies in atomic modes based on Formula \ref{inverse_Bogoliubov_nk},
while diamonds are numerical results.
The difference between analitical and numerical results comes from neglecting the finite width of the quasiparticle spectra in our analitical calculations.
} \label{ekwip_quasi}
\end{figure}

The excitation spectrum of quasiparticles of the system with the condensate in $\textbf{k}_c$ mode has been plotted in Figure \ref{excitation_spectra}. 
It can be well described by (see below):
\begin{equation}\label{excitations_eqaution}
  \epsilon_{\textbf{k}}=\mu+\epsilon \sqrt{\omega(\textbf{k}-\textbf{k}_c)^2-(gn_c)^2}+
  \epsilon (\textbf{k}-\textbf{k}_c) \textbf{k}_c.  
\end{equation}
Relative energies of quasiparticles moving faster than the condensate are shifted upwards, 
and moving slower than the condensate are shifted downwards with respect to the stationary state, 
due to the Doppler-like shift (the last term in Formula \ref{excitations_eqaution}).
Note, that numerically obtained spectra of quasiparticles have finite width, which can be interpreted as the their finite life-time \cite{bpgrJPB2004}.
Still, the quasiparticles obey the equipartition of energies.
Their populations $n_{\delta_{\textbf{k}}}=\delta_{\textbf{k}}^*\delta_{\textbf{k}}$ and energies $\epsilon_{\textbf{k}}$ fulfill a formula:
\begin{equation}\label{equipartition}
  n_{\delta_{\textbf{k}}} (\epsilon_{\textbf{k}} - \mu) = k_B T,
\end{equation}
as can be seen in Figure \ref{ekwip_quasi}a.
This indicates that the system undergoing a superfluid flow is in a thermal equilibrium. 
However, the equipartition occurs not in the condensate nor the thermal cloud's frame, but in the center-of-mass frame instead.

Please confront the relation \ref{equipartition} with a similar one obtained for atomic populations $n_{\textbf{k}}=\alpha_{\textbf{k}}^*\alpha_{\textbf{k}}$ 
(Fig. \ref{ekwip_quasi}b), which do not obey the equipartition.
For phonon modes, where quasiparticles differ significantly from atoms, their populations are smaller than relative atomic populations.

For starting $n_c=8\%$ the critical velocity corresponds to a momentum lower than the momentum of the $\textbf{k}=(1,0,0)$ mode and thus 
any motion of the condensate with respect to the thermal cloud results in a friction.
 
Within our accuracy the observed critical velocity is identical with the one obtained from Landau's criterion \cite{lJPMoscow1941_superfluidity} applied 
to the excitation spectrum in the stationary case (Formula \ref{excitations_stationary}).
Note, however, that even for subcritical velocities there exist thermal atoms that move supercritically with respect to the condensate.
Applying the Landau criterion to each quasiparticle separately would always result in a net drag force, 
as there is more supercritical atoms moving opposite to the condensate than supercritical atoms moving along with the condensate.

The manifestation of superfluidity in our model is the existence of many thermal equilibria, all varying in the relative velocity between the condensate and the thermal cloud. 
Such a velocity is an additional control parameter.
Please note that we have not yet presented any quantitative measure of the amount of the superfluid fraction.
In the next section we present analytical discussion and derive the formula for the superfluid population.

\section{Analitical Results}\label{Analitical Results}

We consider here Bogoliubov excitations at non-zero temperature.
The starting point of our investigation will be Eq. \ref{Heisenberg} for the Bose field operator $\hat{\Psi}$.
Let us choose a reference frame in which the condesate is in $k_c$ mode.
Changing $k_c$ will be interpreted for a while as changing the reference frame.
We decompose $\hat{\Psi}$ into a basis of plane waves:
\begin{equation}\label{relabeling_modes}
\hat{\Psi}=\frac{1}{\sqrt{L^3}} \sum_{\textbf{k}} e^{2\pi i (\textbf{k}+\textbf{k}_c)\textbf{r}/L} \hat{a}^{\prime}_{\textbf{k}}(t)  
\end{equation}
with indices chosen in such a way that $\hat{a}^{\prime}_0$ represents the anihilation operator of the condensate mode.
The equation for $\hat{a}^{\prime}_{\textbf{k}}$ is:
\begin{equation}\label{PlaneOperator}
\begin{split}
  \partial_t \hat{a}^{\prime}_{\textbf{k}}(t)=-i\frac{2\pi^2\hbar}{mL^2}\ (\textbf{k}+\textbf{k}_c)^
  2\ \hat{a}^{\prime}_{\textbf{k}}(t)-
  i\frac{4\pi\hbar a}{mL^3}\\
  \sum_{\textbf{k}_1,\ \textbf{k}_2} 
  \hat{a}^{\prime \dagger}_{\textbf{k}_1}\hat{a}^{\prime}_{\textbf{k}_2}\hat{a}^{\prime}_{\textbf{k}+\textbf{k}_1-\textbf{k}_2}.
\end{split}
\end{equation}

It is a non-linear operator equation.
We linearize it in a way similar to the Bogoliubov-Popov approximation.
For certain elements in the sum in Eq. \ref{PlaneOperator} we group two operators in pairs and replace them with apporpriate mean values.
And thus, terms $a^{\dagger}_{\textbf{k}} a_{\textbf{k}}$ are replaced with $<a^{\dagger}_{\textbf{k}} a_{\textbf{k}}>=N n_{\textbf{k}}$,
which do not depend on time for a stationary state.
$a^{\prime}_0 a^{\prime}_0$ is also replaced by $N n_c e^{-2i\mu t}$, 
where the chemical potential $\mu$ is obtained from the equation for $\hat{a}^{\prime}_0$.
The remaining terms in the summation, which are still non-linear in operators, are neglected.
Results from \cite{bpgrJPB2004} suggest that this approximates the full dynamics rather well.
The neglected terms become important at temperatures close to the critical temperature, limiting the scope of our analysis.

The approximate equation for $\hat{a}^{\prime}_0$ is:
\begin{equation}\label{equation_for_a0}
  i \hbar \partial_t \hat{a}^{\prime}_0 (t)=\epsilon \frac{\textbf{k}_c^2}{2}\ \hat{a}^{\prime}_{0}+
  \epsilon (gn_c + 2gn_T) \hat{a}^{\prime}_0,
\end{equation}
where $n_T$ is the thermal fraction, defined as $n_T=1-n_c$.
It gives the leading contribution to the frequency of the condensate mode $\mu=gn_c+2gn_T+\frac{k_c^2}{2}$, 
which is in agreement with results from the Popov approximation \cite{sp2003book_BEC}.
Corrections to $\mu$ have been discussed in \cite{bpgrJPB2004}.
For $\textbf{k} \neq 0$ we obtain:
\begin{equation}\label{equation_for_ak}
  i \hbar \partial_t \hat{a}^{\prime}_{\textbf{k}} (t)=\epsilon \left( \frac{(\textbf{k}+\textbf{k}_c)^2}{2}\ +
  2 g \right) \hat{a}^{\prime}_{\textbf{k}} + \epsilon gn_c e^{-2i\mu t} \hat{a}^{\prime *}_{-\textbf{k}}.
\end{equation}
The solutions of these equations are:
\begin{equation}\label{soltions_ak}
\begin{split}
  \hat{a}^{\prime}_{\textbf{k}} (t)= \hat{a}^{\prime}_{\textbf{k}}(0)
  \frac{gn_c\omega(\textbf{k})\ e^{-i \mu t}}{2\sqrt{\omega(\textbf{k})^2-(gn_c)^2}} 
  \left( e^{-i \epsilon^+_{\textbf{k}}t}+e^{-i \epsilon^-_{\textbf{k}}t}\right)+\\
  \hat{a}^{\prime}_{\textbf{k}}(0)\ \frac{gn_c}{2}\ e^{-i \mu t} \left( e^{-i \epsilon^+_{\textbf{k}}t}-e^{-i \epsilon^-_{\textbf{k}}t}\right)+\\
  \hat{a}^{\prime \dagger}_{-\textbf{k}}(0) 
  \frac{gn_c\ e^{-i \mu t}}{2\sqrt{\omega(\textbf{k})^2-(gn_c)^2}} \left( e^{-i \epsilon^+_{\textbf{k}}t}-e^{-i \epsilon^-_{\textbf{k}}t} \right),
\end{split}
\end{equation}
where:
\begin{equation}\label{energies_+-}
\begin{array}{ll}
 \epsilon^+_{\textbf{k}}= \epsilon \sqrt{\omega(\textbf{k})^2-(gn_c)^2}+\epsilon \textbf{k} \textbf{k}_c\\
 \epsilon^-_{\textbf{k}}= -\epsilon \sqrt{\omega(\textbf{k})^2-(gn_c)^2}+\epsilon \textbf{k} \textbf{k}_c.
\end{array}
\end{equation}

Now we perform Bogoliubov transformation.
We introduce new bosonic anihilation operators:
\begin{equation}\label{quaisparticles}
  \hat{\delta}_{\textbf{k}}=U_{\textbf{k}} \hat{a}^{\prime}_{\textbf{k}}+
  V_{\textbf{k}}\hat{a}^{\prime \dagger}_{-\textbf{k}} e^{-2i (\mu+\epsilon\textbf{k}\textbf{k}_c) t}.
\end{equation}
which destroy a Bogoliubov excitation in mode $\textbf{k}$.
The commutation relation $[\hat{\delta}_{\textbf{k}},\ \hat{\delta}^{\dagger}_{\textbf{k}^\prime}]=\delta_{\textbf{k},\textbf{k}^\prime}$
implies that $|U|^2-|V|^2=1$.
We look for normal modes of the system (which oscillate with a single frequency only),
which are linear combinations of solutions \ref{soltions_ak}.
The resulting Bogoliubov quasiparticles evolve with the frequency $\mu+\epsilon^+_{\textbf{k}}$:
\begin{equation}\label{quasiparticle_evolution}
\hat{\delta}_{\textbf{k}}(t)=\hat{\delta}_{\textbf{k}}(0) e^{-i (\mu+\epsilon^+_{\textbf{k}}) t}
\end{equation}
and the coefficients $U_{\textbf{k}}$ and $V_{\textbf{k}}$ are given by:
\begin{equation}\label{quasiparticle_UV}
\left\{ \begin{array}{ll}
  U_{\textbf{k}}^2=\frac{\left( \omega(\textbf{k})+\sqrt{\omega(\textbf{k})^2-(gn_c)^2} \right)^2}
  {\left( \omega(\textbf{k})+\sqrt{\omega(\textbf{k})^2-(gn_c)^2}\right)^2-\left( gn_c\right)^2}\\
  V_{\textbf{k}}^2=\frac{(gn_c)^2}
  {\left(\omega(\textbf{k})+\sqrt{\omega(\textbf{k})^2-(gn_c)^2}\right)^2-\left( gn_c\right)^2}
\end{array} \right.
\end{equation}
Energies $\epsilon^+_{\textbf{k}}$ form the excitation spectrum of the system. 
They agree with values obtained from the numerical simulations.
However the calculated value of the chemical potential $\mu$ agrees rather with the Formula $14$ of \cite{bpgrJPB2004} 
than with the simplified Equation \ref{equation_for_a0}.
The difference, however, shifts the whole energy spectrum only.
Note, that this simplified approach ignores the finite width (which determines the life-time) of the quasiparticle spectra,
which we observe in our numerical simulations.

We can also perform an inverse Bogoliubov transofmation.
It is:
\begin{equation}\label{inverse_Bogoliubov}
  \hat{a}^{\prime}_{\textbf{k}} = U_{\textbf{k}} \hat{\delta}_{\textbf{k}} - V_{\textbf{k}} \hat{\delta}^{\dagger}_{-\textbf{k}} 
  e^{-2i (\mu+\epsilon\textbf{k}\textbf{k}_c) t}
\end{equation}
and the resulting atomic populations are:
\begin{equation}\label{inverse_Bogoliubov_nk_quant}
N n_{\textbf{k}}=<\hat{a}^{\prime\dagger}_{\textbf{k}} \hat{a}^{\prime}_{\textbf{k}}>=
U_{\textbf{k}}^2 <\hat{\delta}^{\dagger}_{\textbf{k}}\hat{\delta}_{\textbf{k}}>+
V_{\textbf{k}}^2 <\hat{\delta}_{-\textbf{k}}\hat{\delta}^{\dagger}_{-\textbf{k}}>.
\end{equation}
Note that $U_{\textbf{k}}=U_{-\textbf{k}}$, $V_{\textbf{k}}=V_{-\textbf{k}}$.

The resulting Bogoliubov quasiparticles $\delta_{\textbf{k}}$ have momenta 
$p_{\delta_{\textbf{k}}}=\hbar (\textbf{k}+\textbf{k}_c)$
and their coefficients $U_{\textbf{k}}$ and $V_{\textbf{k}}$ are independent of the choice of the reference frame ($\textbf{k}_c$).
Thus they transform under Galilean transform as particles of mass m, similarly to atoms in our system. 

Let us now switch to the semi-classical model again, by substituting $c$-numbers for the operators
($\hat{a}^{\prime}_{\textbf{k}} \rightarrow \alpha_{\textbf{k}}$ 
and $\hat{\delta}_{\textbf{k}} \rightarrow \delta_{\textbf{k}}$).
By doing this we neglect the antinormal ordering of $\hat{\delta}_{\textbf{k}}$ in Formula \ref{inverse_Bogoliubov_nk_quant},
which accounts for the quantum depletion.
On the other hand we can relate the results to our numerical simulations, 
allowing us to calculate the amount of superfluid fraction.

In this semi-classical approach Formula \ref{inverse_Bogoliubov_nk_quant} takes the form:
\begin{equation}\label{inverse_Bogoliubov_nk}
n_{\textbf{k}}=
U_{\textbf{k}}^2 n_{\delta_{\textbf{k}}}+V_{\textbf{k}}^2 n_{\delta_{-\textbf{k}}},
\end{equation}
where $n_{\delta_{\textbf{k}}}=\delta^*_{\textbf{k}}\delta_{\textbf{k}}$ are quasiparticle populations.
Assuming we know quasiparticle populations we can calculate populations of atomic modes with this formula. 
They are greater than the corresponding quasiparticle populations due to $U_{\textbf{k}}^2\ge 1$.
These analitical calculations agree with our numerical results
(see Fig. \ref{ekwip_quasi}b for comparison of atomic populations).

In order to identify the superfluid part we consider the total momentum $P$ of the system.
It can be naturally expressed as a sum over momenta of individual atoms:
\begin{equation}\label{momentum_atomic}
P=\hbar N\left( n_c \textbf{k}_c+\sum_{\textbf{k} \neq 0} n_{\textbf{k}} (\textbf{k}+\textbf{k}_c) \right)=
P_c+P_T,
\end{equation}
where atoms contribute either to the momentum of the condensate $P_c$ or
to the momentum of the thermal cloud $P_T$ (for $\textbf{k} \neq 0$).

On the other hand quasiparticles have a well-defined momenta too and they are also the observed excitations
that can be measured experimentally \cite{jPRL1996_experimental_quasiparticles,mPRL1996_experimental_quasiparticles,vxrakPRL2002_experimental_quasiparticles}.
Their agregate momentum is
$P_{\delta}=N \hbar \sum_{\textbf{k} \neq 0} n_{\delta_{\textbf{k}}} (\textbf{k}+\textbf{k}_c)$
and we can express the total momentum $P$:
\begin{equation}\label{momentum_quasiparticles}
P=P_{\delta_{\textbf{k}}} + P_{SF}.
\end{equation}
An additional momentum $P_{SF}$ appears here to account for the conservation of momentum. 
Because $P_{\delta}$ and $P$ are physically observable $P_{SF}$ should have some physical meaning.
Applying the inverse Bogoliubov transformation (\ref{inverse_Bogoliubov_nk}) to atomic populations 
and using the relations for $U_{\textbf{k}}$ and $V_{\textbf{k}}$ one can show that:
\begin{equation}\label{theraml_quasi_relation}
\begin{split}
n_T=n_{\delta} + 2 \sum_{\textbf{k} \neq 0} n_{\delta_{\textbf{k}}} V_{\textbf{k}}^2,
\end{split}
\end{equation}
This implies that:
\begin{equation}\label{momentum_superfluid}
P_{SF}= N \hbar (n_c+n_T-n_{\delta}) \textbf{k}_c,
\end{equation}
where $n_{\delta}=\sum_{\textbf{k} \neq 0} n_{\delta_{\textbf{k}}}$ is the total number of quasiparticles.
Only in the condensate reference frame this momentum is zero.
It appears to come from a fraction $n_{SF}=n_c+n_T-n_{\delta}=1-n_{\delta}$
moving with the condensate velocity.
This mass represents condensed atoms along with a part of thermal cloud "freezed out" by the interaction.

We have decomposed the system into two parts - one composed of excitations of the system and another one moving with the condensate speed.
Applying the Landau criterion to such a system we conclude 
that the $n_{SF}=1-n_{\delta}$ is indeed the superfluid fraction, as indicated by the subscript we used,
and Bogoliubov quasiparticles form the normal fraction in the system ($n_{\delta}$).

In other words the superfluid fraction is a sum of the condensate and atoms which do not take part in forming quasiparticles 
(which are thermally distributed).
Thus atoms occupying highly excited modes do not contribute to the superfluid part, 
as for these modes quasiparticles are almost the same as atomic modes.
Note that the superfluid fraction adds up with quasiparticles to the total number of atoms in the system, 
even though Bogoliubov quasiparticles do not obey the conservation of their number.

Due to the bosonic nature of Bogoliubov quasiparticles the superfluid fraction is greater than the condensate fraction and the difference increases with the interaction strength $g$. 
In our numerical simulations it is for example $n_{SF}= 60.4\%$ compared to $n_c= 53.4\%$ 
for relative velocity $v=0.82 c$ between fractions.

We can apply the same reasoning without restricting ourselves to the semi-classical description of atoms and quasiparticles.
The same scheme applied to a system with quantum corrections taken into consideration shows that even at zero temperature the superfluid fraction
 is $100\%$ even though the condensed fraction is smaller than that.
Because the quantum depletion originates from the reversed order of $\delta_{\textbf{k}}$ in the inverse Bogoliubov transform,
the formula for superfluid population takes the form:

\begin{equation}\label{superfluid_quantum}
\begin{split}
N n_{SF}=N n_c+
\sum_{\textbf{k} \neq 0} \left( U_{\textbf{k}}^2 <\hat{\delta}^{\dagger}_{\textbf{k}} \hat{\delta}_{\textbf{k}}> + 
V_{\textbf{k}}^2 <\hat{\delta}_{-\textbf{k}} \hat{\delta}^{\dagger}_{-\textbf{k}}>\right)-\\
\sum_{\textbf{k} \neq 0} <\hat{\delta}^{\dagger}_{\textbf{k}} \hat{\delta}_{\textbf{k}}>\\
=N n_c+N n_T-N n_{\delta}+N_{QD},
\end{split}
\end{equation}
whereas the total number of atoms is given by:
\begin{equation}\label{total_atoms_quantum}
\begin{split}
N=N n_c+\sum_{\textbf{k} \neq 0} \left( U_{\textbf{k}}^2 <\hat{\delta}^{\dagger}_{\textbf{k}} \hat{\delta}_{\textbf{k}}> + 
V_{\textbf{k}}^2 <\hat{\delta}_{-\textbf{k}} \hat{\delta}^{\dagger}_{-\textbf{k}}> \right)\\
=N n_c+N n_T+N_{QD}.
\end{split}
\end{equation}
In these formulas $N n_{\delta}=\sum_{\textbf{k} \neq 0} <\hat{\delta}^{\dagger}_{\textbf{k}} \hat{\delta}_{\textbf{k}}>$,
$N n_T=N n_{\delta} + 2 \sum_{\textbf{k} \neq 0} \left( <\hat{\delta}^{\dagger}_{\textbf{k}} \hat{\delta}_{\textbf{k}}> V_{\textbf{k}}^2 \right)$
and the quantum depletion $N_{QD}=\sum_{\textbf{k} \neq 0} V_{\textbf{k}}^2$ is accounted separately.
At zero temperature, when $<\hat{\delta}^{\dagger}_{\textbf{k}}\hat{\delta}_{\textbf{k}}>=0$,
both $n_T$ and $n_{\delta}$ vanish and the superfluid population equals the total number of atoms in the system.

Even though such a basic model suffers from a divergence in quantum depletion terms resulting from the zero range potential,
it clearly indicates that the quantum depletion belongs to the superfluid fraction.

\section{Conclusions}\label{Conclusions}

In this paper we have presented a model of superfluidity of a weakly-interacting Bose gas within 
the classical fields approximation.
In the direct numerical simulation of the dynamics we have shown that below the Landau critical velocity 
$c=\frac{2 \pi \hbar}{mL} \sqrt{gn_c}$
the relative flow of the condensate with respect to the thermal cloud is superfluid 
and that the resulting state is in a thermal equilibrium. 
The relative velocity of this superflow is an additional control parameter characterizing the thermal equilibrium.

We have obtained the excitation spectrum of the system and we have analyzed in detail properties of Bogoliubov quasiparticles, 
obtaining an agreement between numerical and analitical calculations. 

The superfluid fraction has been identified from the momentum transformation rules and an application of the Landau criterion.
It turns out to be the condensate plus atoms from the excited modes which do not participate 
in the collective excitations of the system.
These atoms can be viewed as bound to the condensate by the interactions.
The results provide analitical formula for a difference between superfluid and condensed parts 
at any temperature and within a consistent and non-perturbative description.
They also show that one does not require a strongly-interacting Helium-II, 
with its roton part of the spectrum and essential many-body interactions, to make this difference significant.

In our approach it is also evident that quantum depletion occuring in a weakly-interacting Bose gas is also a part of the superfluid fraction, 
so that even at $0K$ there is a difference between superfluid and condensed fractions - again similarly to He-II. 

The temperature dependent Bogoliubov-Popov spectrum, together with our approach to superfluidity
can form a basis for equilibrium statistical physics of interacting Bose gas at non-zero temperature.
Such a study will be presented elsewhere \cite{future_zgr_GCEquasi}.

\begin{acknowledgments}

We thank P. Navez for providing inspiration for the present paper.
The authors acknowledge support of the Polish Ministry of Scientific Research and Information Technology
under Grant "Quantum Information and Quantum Engineering" No. PBZ-MIN-008/P03/2003. 
The results have been obtained using computers at the Interdisciplinary Centre for
Mathematical and Computational Modeling of Warsaw University.

\end{acknowledgments}

\end{document}